\documentclass[conference]{IEEEtran}

\usepackage[T1]{fontenc}
\usepackage[utf8]{inputenc}
\usepackage{graphicx}
\usepackage{booktabs}
\usepackage{array}
\usepackage{xcolor}
\usepackage{hyperref}
\usepackage{xurl}
\usepackage{amsmath}
\usepackage{microtype}
\usepackage{cite}
\usepackage{balance}
\usepackage{float}
\hypersetup{colorlinks=true, linkcolor=black, urlcolor=blue, citecolor=black}

\newcolumntype{C}[1]{>{\centering\arraybackslash}p{#1}}
\newcolumntype{L}[1]{>{\raggedright\arraybackslash}p{#1}}

\begin{document}

\title{Shards on a Shoestring: Empirical Characterization of NEAR Protocol Nightshade Sharding on Commodity Hardware}

\author{
  \IEEEauthorblockN{Sohini Sahukar}
  \IEEEauthorblockA{
    Illinois Institute of Technology\\
    Chicago, IL, USA\\
    ssahukar@hawk.illinoistech.edu
  }
  \and
  \IEEEauthorblockN{Om Amit Gandhi}
  \IEEEauthorblockA{
    Illinois Institute of Technology\\
    Chicago, IL, USA\\
    ogandhi1@hawk.iit.edu
  }
  \and
  \IEEEauthorblockN{Ioan Raicu}
  \IEEEauthorblockA{
    Illinois Institute of Technology\\
    Chicago, IL, USA\\
    iraicu@cs.iit.edu
  }
}

\maketitle

\begin{abstract}
NEAR Protocol's Nightshade architecture targets one million transactions
per second (TPS) through horizontal sharding of both state and
computation. Published benchmarks were produced on expensive Google Cloud
Platform infrastructure costing approximately \$700 per hour, leaving a
significant reproducibility gap for academic research. We present the
first independent empirical characterization of NEAR Nightshade sharding
on commodity hardware: a Chameleon Cloud bare-metal node with 48
hyperthreaded Intel Xeon cores, 128\,GB RAM, and HDD storage at
80--100\,MB/s. We systematically sweep shard count from $N{=}1$ to
$N{=}24$, measuring aggregate TPS, per-shard TPS, block time, BFT
finality, memory, and disk I/O. We identify three distinct bottleneck
regimes: L3 cache pressure at low $N$, witness gossip pipeline saturation
at mid $N$, and coherence collapse at high $N$. A key unexpected finding
is that HDD write latency acts as implicit flow control for the witness
gossip pipeline. Removing it via RAM-backed tmpfs causes complete chain
stall at $N{=}16$, with a 29$\times$ spike in orphan witness rate at 47\%
CPU utilization. Aggregate TPS peaks at $N{=}8$ (+40\% over $N{=}1$) then
reverses, with per-shard TPS collapsing 23$\times$ by $N{=}24$. Our
dataset provides the first commodity-hardware calibration baseline for the
companion SimPy sharding simulator.
\end{abstract}

\begin{IEEEkeywords}
blockchain sharding, NEAR Protocol, Nightshade, empirical evaluation,
Chameleon Cloud, reproducibility, distributed systems
\end{IEEEkeywords}

\section{Introduction}

Blockchain scalability is one of the central unsolved problems in
distributed systems. Bitcoin processes approximately 6 transactions per
second. Visa handles around 24,000. NEAR Protocol, through its Nightshade
sharding architecture~\cite{nightshade}, targets one million TPS. This
claim would fundamentally change what is possible in decentralized
application infrastructure if it can be independently verified and
reproduced.

NEAR's Nightshade design shards both state and computation. Each shard is
assigned a dedicated \emph{chunk producer} responsible for executing
transactions and producing \emph{state witnesses}: cryptographic proofs of
valid state transitions. These witnesses are gossiped across the entire
validator set before any block can be finalized, creating an
$\mathcal{O}(N^2)$ communication structure that constrains how the system
scales as shard count grows.

NEAR's own published benchmarks are produced on expensive Google Cloud
Platform infrastructure costing approximately \$700 per hour. These results
are unreproducible by academic research groups without significant funding.
No independent empirical characterization of NEAR sharding on commodity
hardware existed prior to this work.

This paper is the empirical companion to Gandhi and
Raicu~\cite{gandhi2026}, which presents a configurable SimPy-based
discrete-event simulator for evaluating sharded blockchain architectures.
That work predicts a throughput crossover point beyond which coordination
overhead outweighs parallelism gains. Our hardware measurements confirm
and quantify this prediction for the first time on real NEAR Protocol
infrastructure. The core contributions are:

\begin{itemize}
  \item The first independent empirical characterization of NEAR Nightshade
    sharding on commodity hardware (Chameleon Cloud bare-metal).
  \item A systematic shard scaling sweep from $N{=}1$ to $N{=}24$ with
    detailed block time decomposition via Prometheus instrumentation.
  \item Identification and mechanistic explanation of three distinct
    bottleneck regimes.
  \item Discovery that HDD backpressure serves as implicit flow control the
    protocol depends on.
  \item Real-world pipeline calibration data for the companion SimPy
    sharding simulator~\cite{gandhi2026}.
\end{itemize}

\section{Background}

\subsection{NEAR Nightshade Architecture}

NEAR Protocol is a layer-1 proof-of-stake smart contract platform launched
on mainnet in October 2020, co-founded by Alex Skidanov and Illia
Polosukhin. NEAR shards the account namespace: every account belongs to
exactly one shard, determined by a hash of the account name. A NEAR block
is a logical container that aggregates one chunk per shard. A \emph{chunk}
is the physical output of a shard for a given block height, containing: a
list of transactions, incoming cross-shard receipts, the pre-state root,
the post-state root, and the state witness.

In Protocol 84 (the version used in this study), validators maintain a
full local copy of their shard's state in RocksDB and execute transactions
against it to validate witnesses. Each chunk producer gossips its witness
to all $N{-}1$ other validators before the block deadline, producing
$N(N{-}1)$ witness gossip messages per block: 552 at $N{=}24$.

\subsection{Consensus: Doomslug and BFT Finality}

NEAR uses a two-layer consensus architecture. Doomslug provides practical
finality after one round of approval messages ($>$50\% stake online). The
BFT Finality Gadget provides cryptographic irreversibility after two rounds
($>$66\% stake online). Our finality measurements capture BFT finality:
the time for a transaction to achieve full irreversibility. This is why
finality grows from 6.8\,s at $N{=}1$ to 77.5\,s at $N{=}8$.

\subsection{The Reproducibility Gap}

NEAR's published one-million-TPS benchmark used 8 Google Cloud Platform
\texttt{c2d-highcpu-112} instances (112 vCPUs, 224\,GB RAM each) at an
estimated cost exceeding \$700 per hour. The GCP harness~\cite{onemilliontps}
was tightly coupled to Google Cloud APIs and assumed high-performance NVMe
storage. Porting to Chameleon required nine engineering changes: genesis
account baking (replacing RPC-based 50,000-account creation that cascaded
into HDD timeouts), Python 3.10/3.11 compatibility, GCP API removal, RPC
timeout tuning, CPU pinning flags, netem support, measurement window
parameterization, validator seat count correction (\texttt{[1]*N} not
default \texttt{[100]*N}), and RocksDB LOCK file cleanup.

\section{Experimental Setup}

\subsection{Hardware and Configuration}

All experiments ran on a single Chameleon Cloud bare-metal node: Intel
Xeon with 48 hyperthreaded cores at 2.3--3.1\,GHz, 128\,GB RAM, and HDD
storage at 80--100\,MB/s sequential write throughput. The \texttt{neard}
binary was built from commit \texttt{d178e1830} (Protocol 84).

Each experiment spawned $N$ \texttt{neard} validator processes on the
single node, one per shard, sharing all CPU, RAM, and HDD resources.
Genesis accounts (5,000 per shard) were baked into \texttt{genesis.json}
at block zero. Gas limit was set to 30\,Tgas per shard. A critical fix
was setting \texttt{num\_block\_producer\_seats\_per\_shard = [1]*N}: the
default \texttt{[100]*N} with only $N$ validators causes the epoch manager
to scatter validators across all shards, producing near-zero TPS. Each run
used a 30-second ramp-up followed by a 180-second measurement window. CPU
governor was set to performance mode (critical: \texttt{schedutil} caused
significant regressions).

\begin{figure}[H]
  \centering
  \includegraphics[width=\columnwidth]{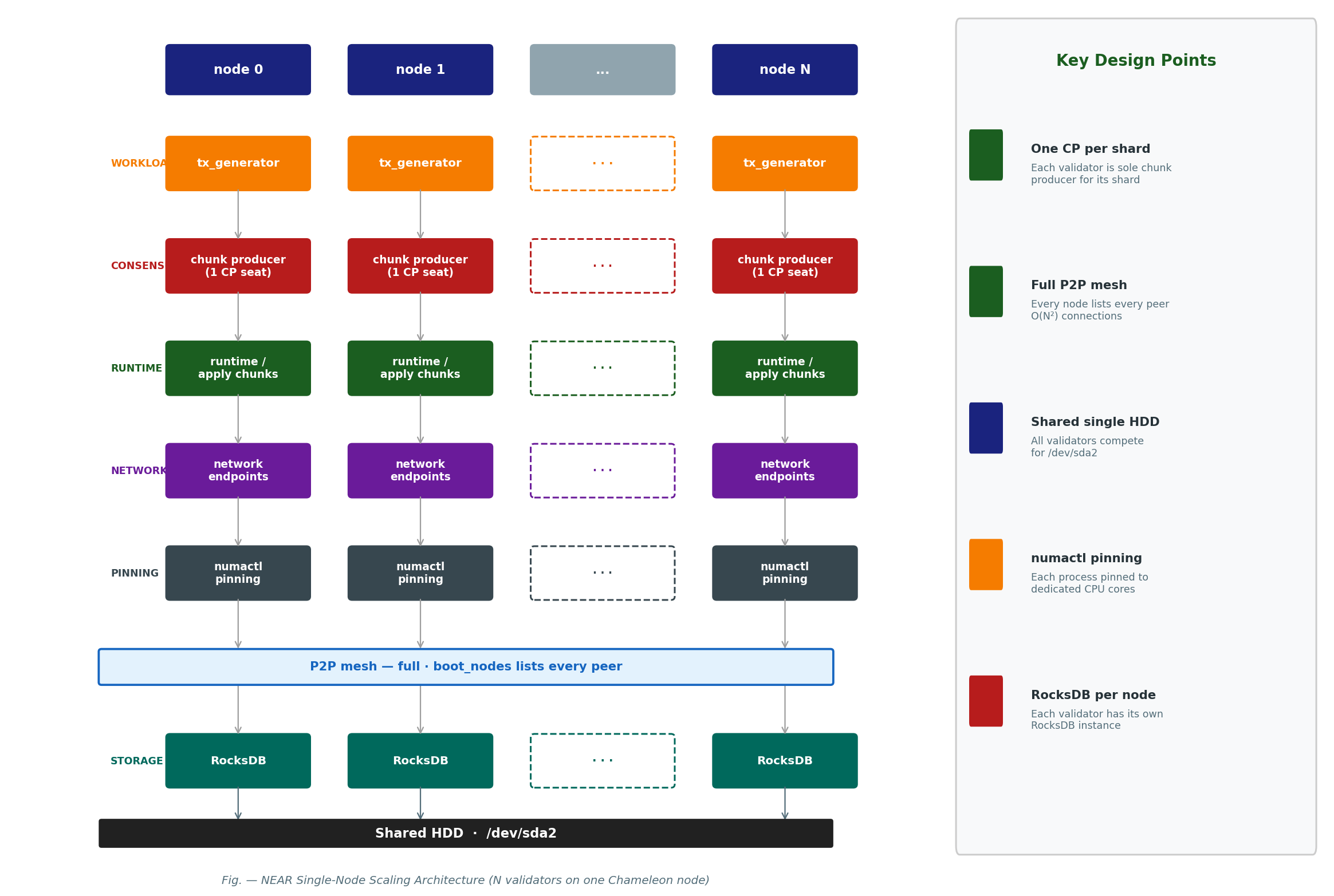}
  \caption{NEAR single-node architecture for Phase 4. $N$ validators
    co-located on one Chameleon node, sharing a single HDD, connected
    via full P2P mesh ($\mathcal{O}(N^2)$ connections), each with an
    independent RocksDB instance.}
  \label{fig:arch}
\end{figure}

Fig.~\ref{fig:arch} shows the single-node architecture used in Phase 4.

\subsection{Monitoring}

A Prometheus scraper polled \texttt{/metrics} every 5 seconds per
validator process, recording transaction counts, block processing time,
chunk produced/skipped totals, delayed receipts count, witness size, RSS,
and disk write throughput. Block-time decomposition used
$\Delta\text{sum}/\Delta\text{count}$ on \texttt{apply\_delay\_ms} across
the steady-state window via \texttt{analyze\_block\_pipeline.py}.

\subsection{Additional Engineering Fixes}

At $N \geq 8$, the epoch manager scrambles validator-to-shard mapping
after the first epoch rotation. The benchmark script was updated to query
RPC post-startup, detect mismatches, and rewrite account submission paths.
After SIGKILL of \texttt{neard} processes, RocksDB LOCK files cause
EAGAIN panics on restart. Fixed by deleting each node's LOCK file before
spawning \texttt{neard} processes.

\section{Results}

\subsection{HDD Scaling Sweep}

\begin{figure}[H]
  \centering
  \includegraphics[width=\columnwidth]{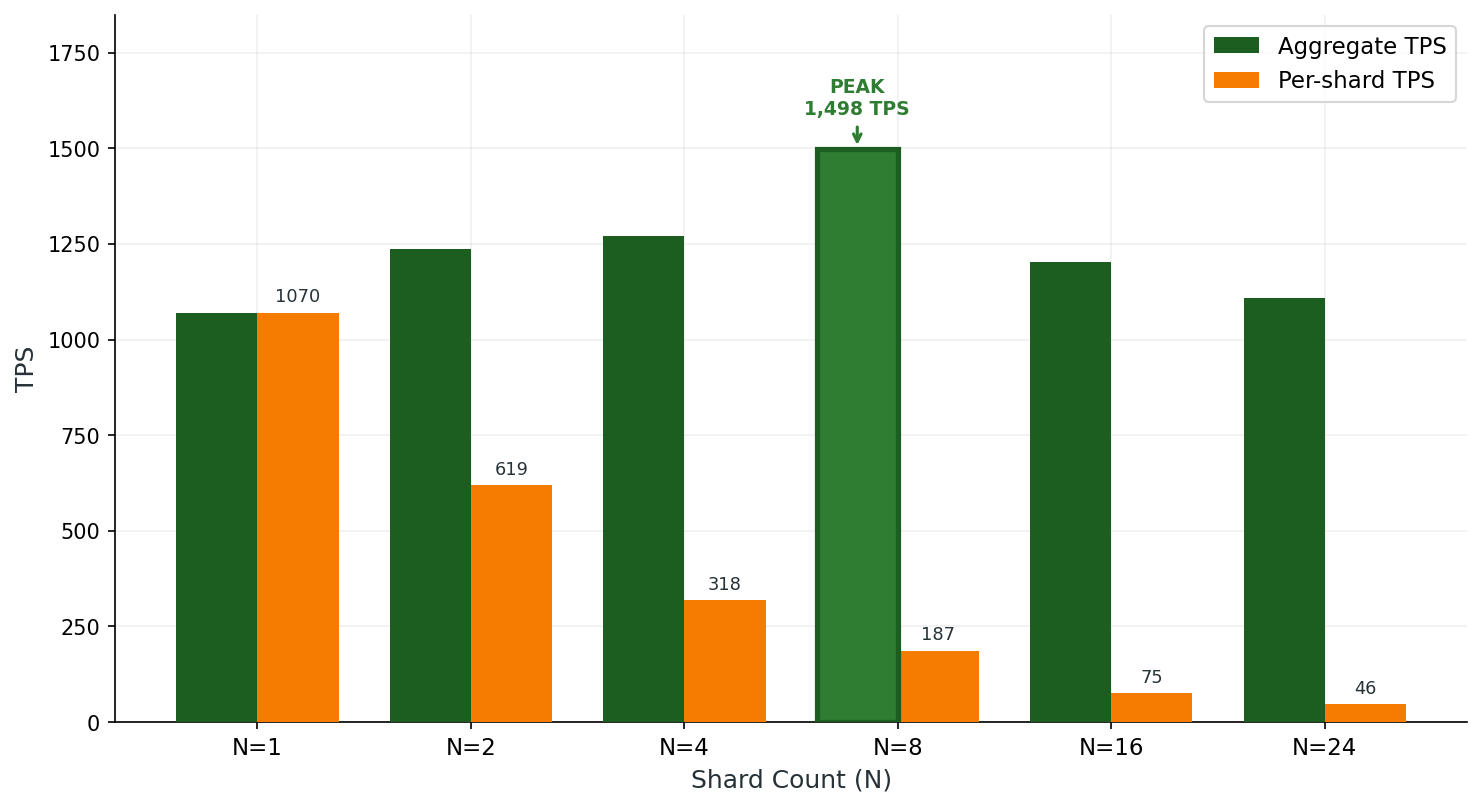}
  \caption{Aggregate TPS (green) and Per-Shard TPS (orange) vs shard
    count $N$. Peak at $N{=}8$. Per-shard TPS collapses 23$\times$ from
    $N{=}1$ to $N{=}24$.}
  \label{fig:tps}
\end{figure}

Table~\ref{tab:hdd} presents the complete HDD scaling results.
Fig.~\ref{fig:tps} shows the throughput picture.

\begin{figure}[H]
  \centering
  \includegraphics[width=\columnwidth]{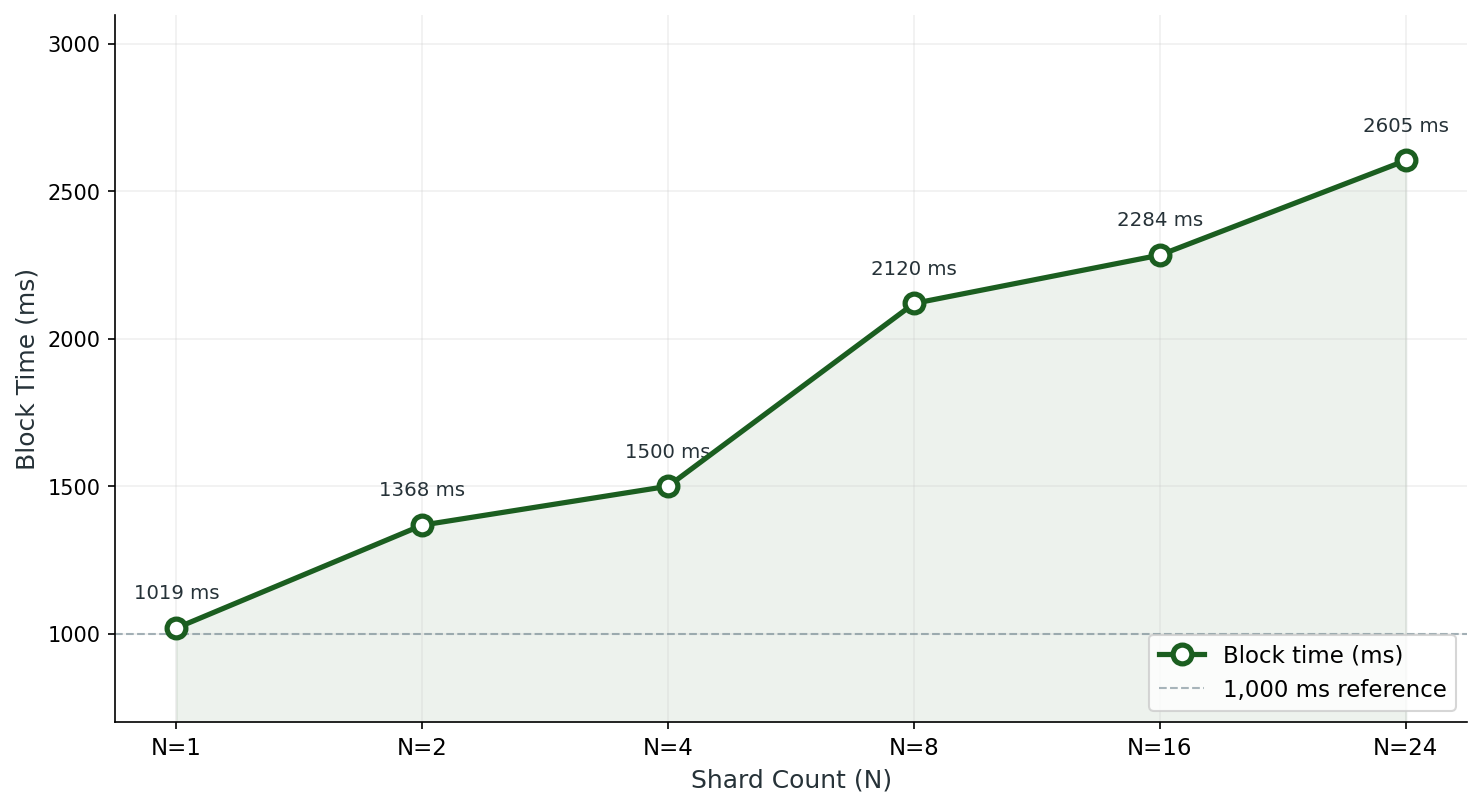}
  \caption{Block time (ms) vs shard count. Monotonic growth from 1,019\,ms
    to 2,605\,ms driven by $\mathcal{O}(N^2)$ BFT approval messages.}
  \label{fig:blocktime}
\end{figure}

Aggregate TPS peaks at $N{=}8$ (1,498 TPS, +40\% over $N{=}1$) then
reverses: 1,204 at $N{=}16$, 1,110 at $N{=}24$. Per-shard TPS collapses
23$\times$ from 1,070 to 46. At $N{=}24$, each shard produces less than
5\% of what a single shard achieves alone. Block time grows monotonically
from 1,019\,ms to 2,605\,ms (Fig.~\ref{fig:blocktime}), driven by
$\mathcal{O}(N^2)$ BFT approval messages: 276 unique validator pairs
exchanging 552 messages per block at $N{=}24$. BFT finality reaches
77.5\,s at the throughput peak ($N{=}8$). Run-to-run CV\% reaches 27.6\%
at $N{=}16$, reflecting genuine threshold behavior as the system operates
near the coherence collapse boundary.

\begin{table}[H]
\centering
\caption{Phase 4 HDD Scaling Results. $\star$ = peak aggregate TPS.
  Red CV\% and RSS values flag anomalous behavior.}
\label{tab:hdd}
\renewcommand{\arraystretch}{1.15}
\begin{tabular}{C{0.4cm} C{0.85cm} C{0.85cm} C{0.9cm} C{0.85cm} C{0.7cm} C{0.5cm}}
\toprule
\textbf{N} & \textbf{Agg TPS} & \textbf{TPS/Sh} & \textbf{Blk Time} &
\textbf{Final.} & \textbf{RSS GB} & \textbf{CV\%} \\
\midrule
1  & 1,070 & 1,070 & 1,019\,ms & 6.8\,s  & 0.86  & 0.0  \\
2  & 1,238 & 619   & 1,368\,ms & 22.9\,s & 1.59  & 5.1  \\
4  & 1,272 & 318   & 1,500\,ms & 31.5\,s & 3.18  & 6.2  \\
\textbf{8$\star$} & \textbf{1,498} & \textbf{187} & \textbf{2,120\,ms} &
  \textbf{77.5\,s} & \textbf{5.59} & \textbf{17.6} \\
16 & 1,204 & 75  & 2,284\,ms & 64.0\,s &
  \textcolor{red}{\textbf{37.80}} & \textcolor{red}{\textbf{27.6}} \\
24 & 1,110 & 46  & 2,605\,ms & 50.2\,s &
  \textcolor{red}{\textbf{19.70}} & \textcolor{red}{\textbf{27.5}} \\
\bottomrule
\end{tabular}
\end{table}

\subsection{Block Time Decomposition}

\begin{figure}[H]
  \centering
  \includegraphics[width=\columnwidth]{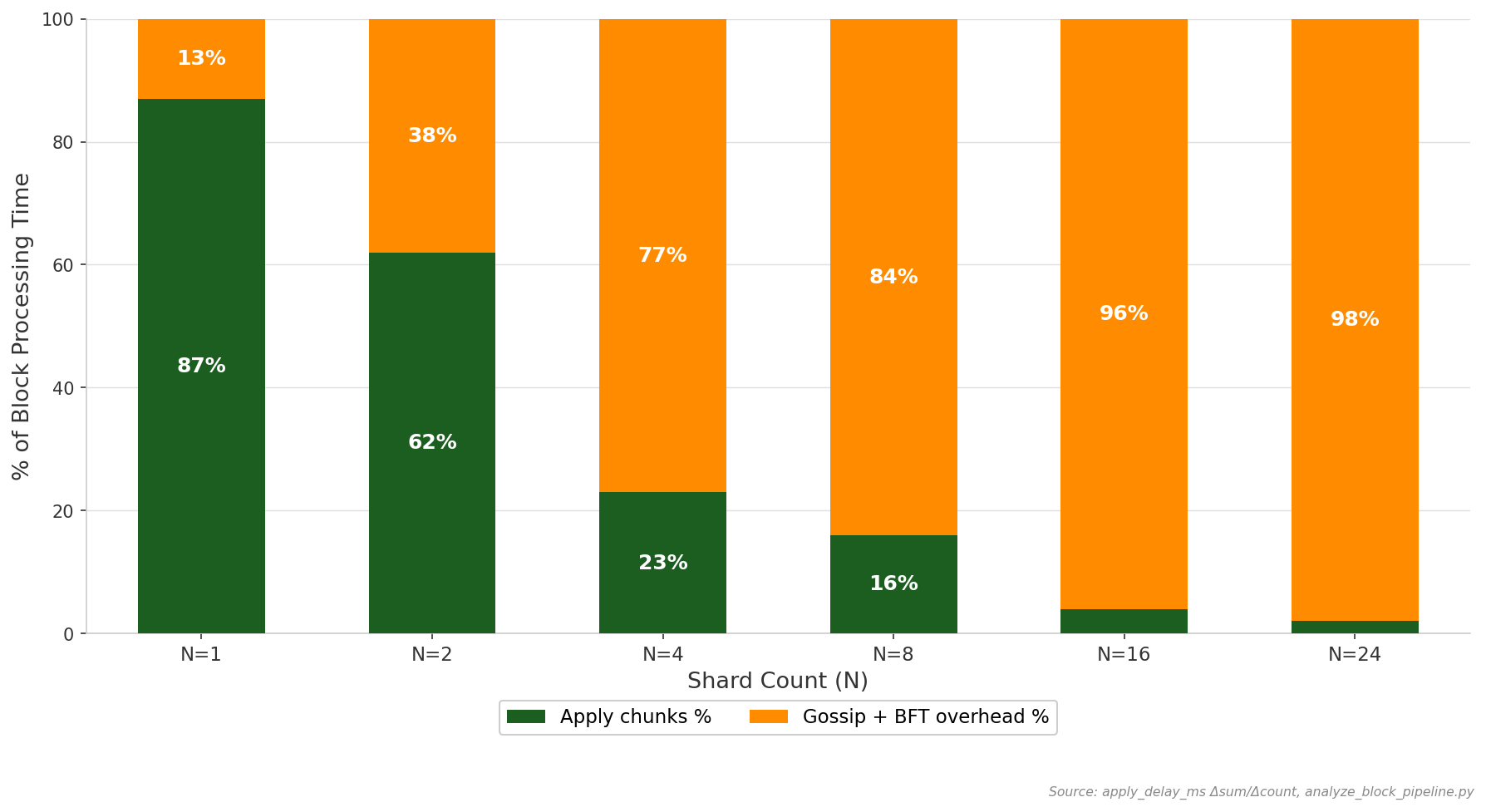}
  \caption{Block time decomposition: \texttt{apply\_chunks}\% (green) vs
    gossip and BFT overhead\% (orange). Apply collapses from 87\% at
    $N{=}1$ to 2\% at $N{=}24$ as gossip overhead dominates.}
  \label{fig:decomp}
\end{figure}

Fig.~\ref{fig:decomp} shows how the dominant cost shifts as $N$ grows,
built from real steady-state Prometheus data
(\texttt{apply\_delay\_ms} $\Delta\text{sum}/\Delta\text{count}$).

At $N{=}1$, \texttt{apply\_chunks} accounts for 1,118\,ms: 87\% of the
1,285\,ms block processing time. By $N{=}4$ this has dropped to 307\,ms
(23\% of 1,307\,ms). By $N{=}16$, apply is only 179\,ms: 4\% of
4,486\,ms total. The 4,307\,ms gap is witness gossip and BFT coordination
overhead. \texttt{produce\_chunk} is always under 100\,ms and is never
the bottleneck at any shard count.

\subsection{Memory and Disk I/O}

\begin{figure}[H]
  \centering
  \includegraphics[width=\columnwidth]{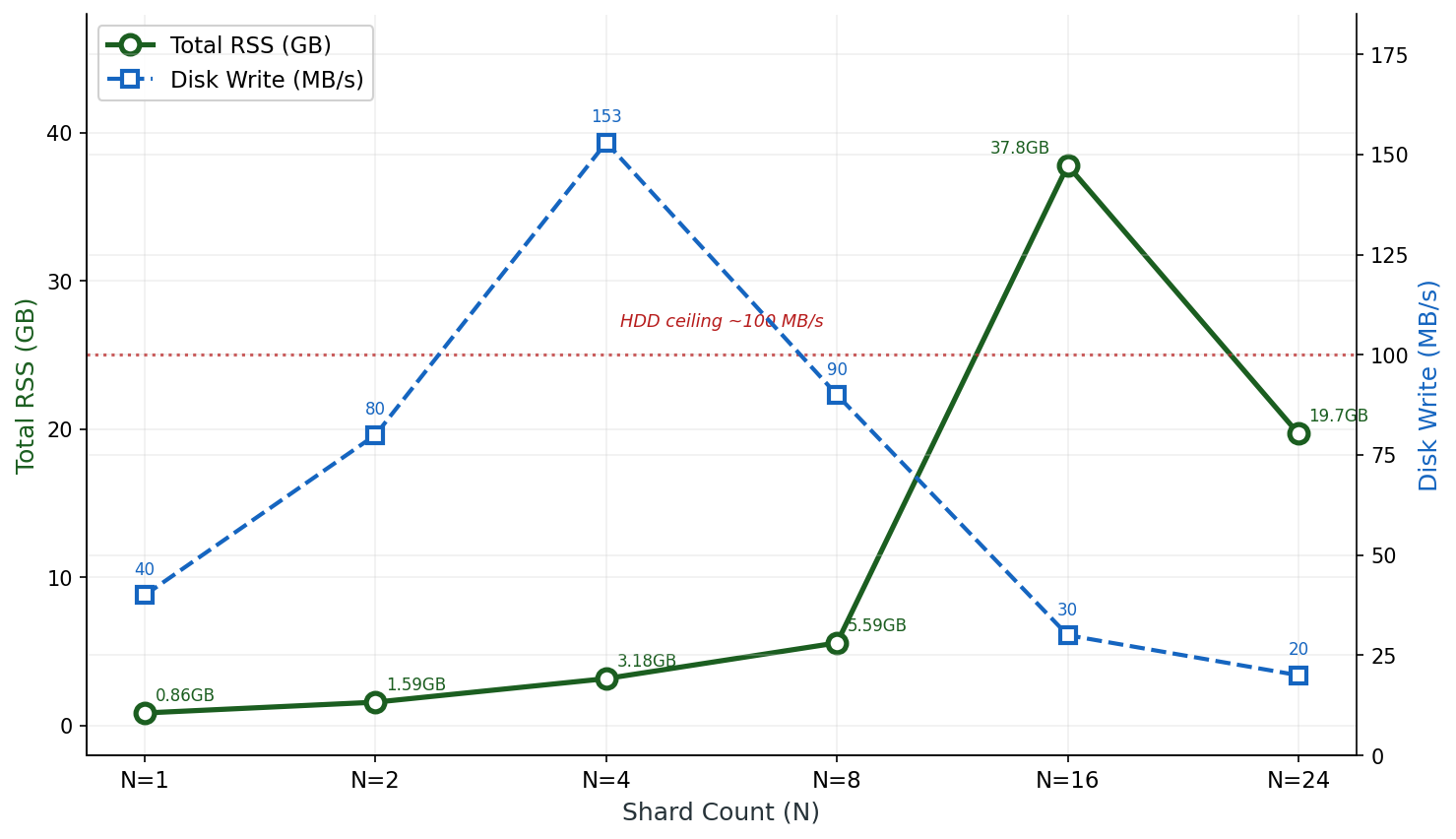}
  \caption{Total RSS (GB, green) and Disk Write throughput (MB/s, blue)
    vs shard count. Red dotted line marks HDD ceiling at
    $\sim$100\,MB/s. Disk peaks at $N{=}4$ (153\,MB/s). RSS spikes
    anomalously to 37.80\,GB at $N{=}16$.}
  \label{fig:rssdisk}
\end{figure}

\begin{figure}[H]
  \centering
  \includegraphics[width=\columnwidth]{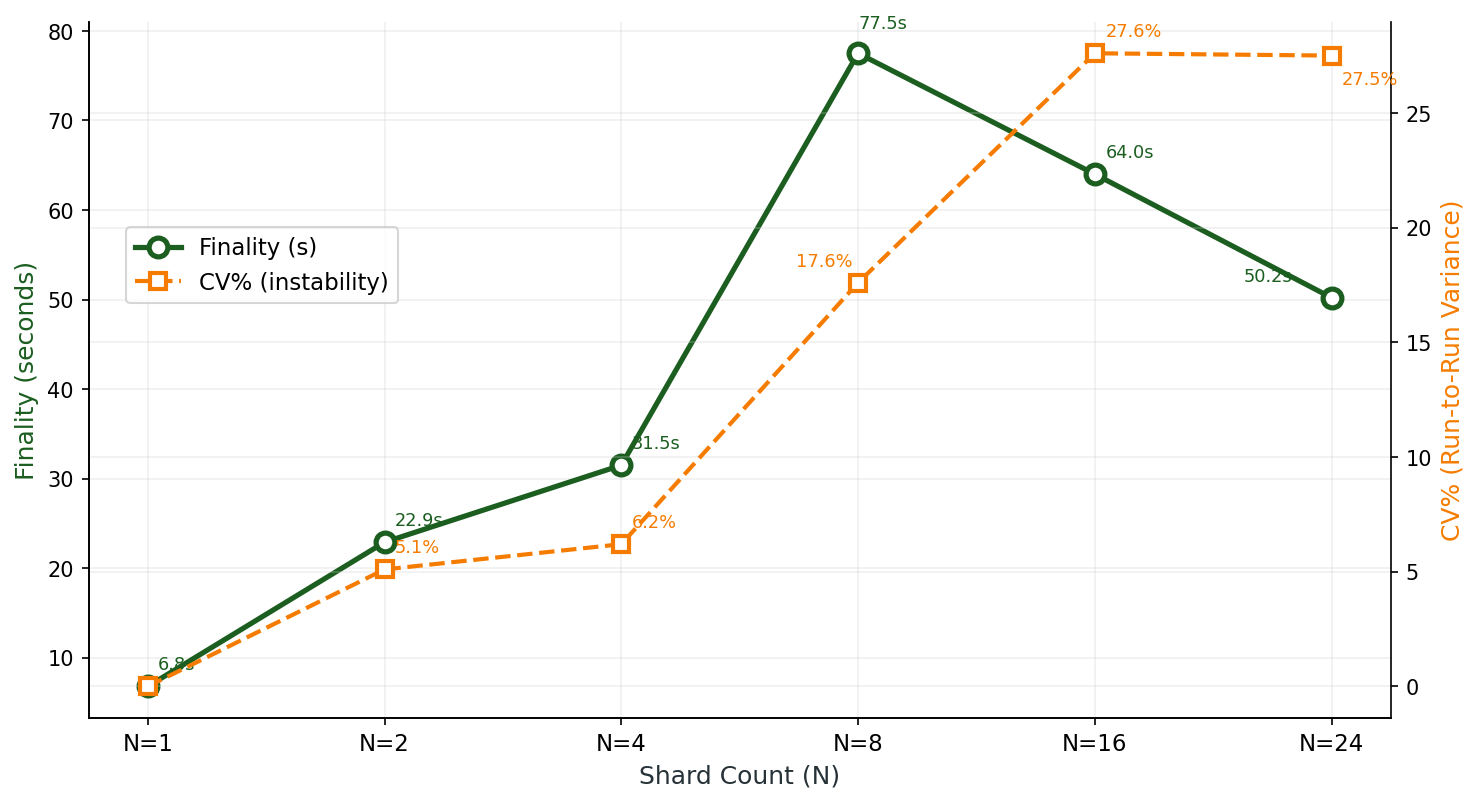}
  \caption{Finality time (s, green) and run-to-run CV\% (orange) vs
    shard count. High CV\% at $N{=}16$ signals instability near the
    coherence collapse threshold.}
  \label{fig:finalitycv}
\end{figure}

Disk writes peak at $N{=}4$ at 153\,MB/s, above the 80--100\,MB/s HDD
ceiling, directly motivating the tmpfs experiment. RSS grows smoothly
from $N{=}1$ through $N{=}8$, then spikes dramatically to 37.80\,GB at
$N{=}16$: a pre-signal of coherence collapse. Finality of 77.5\,s at
$N{=}8$ and CV\%$=$27.6 at $N{=}16$ are practical constraints that matter
as much as raw TPS for production deployments.

\subsection{Chunk Skip Rate and Delayed Receipts}

\begin{figure}[H]
  \centering
  \includegraphics[width=\columnwidth]{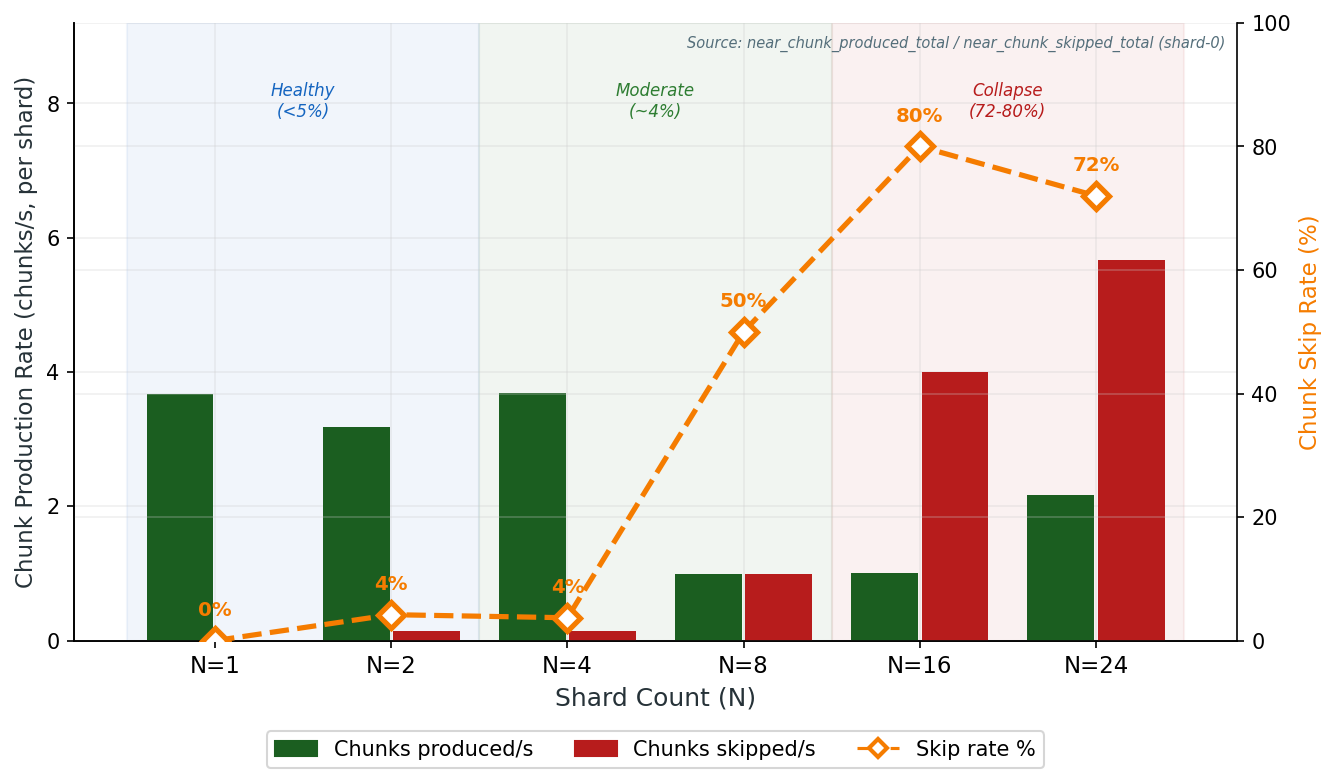}
  \caption{Chunk skip rate per shard. 0\% at $N{=}1$, 50\% at $N{=}8$,
    80\% at $N{=}16$. Measured from \texttt{near\_chunk\_produced\_total}
    and \texttt{near\_chunk\_skipped\_total}.}
  \label{fig:skiprate}
\end{figure}

\begin{figure}[H]
  \centering
  \includegraphics[width=\columnwidth]{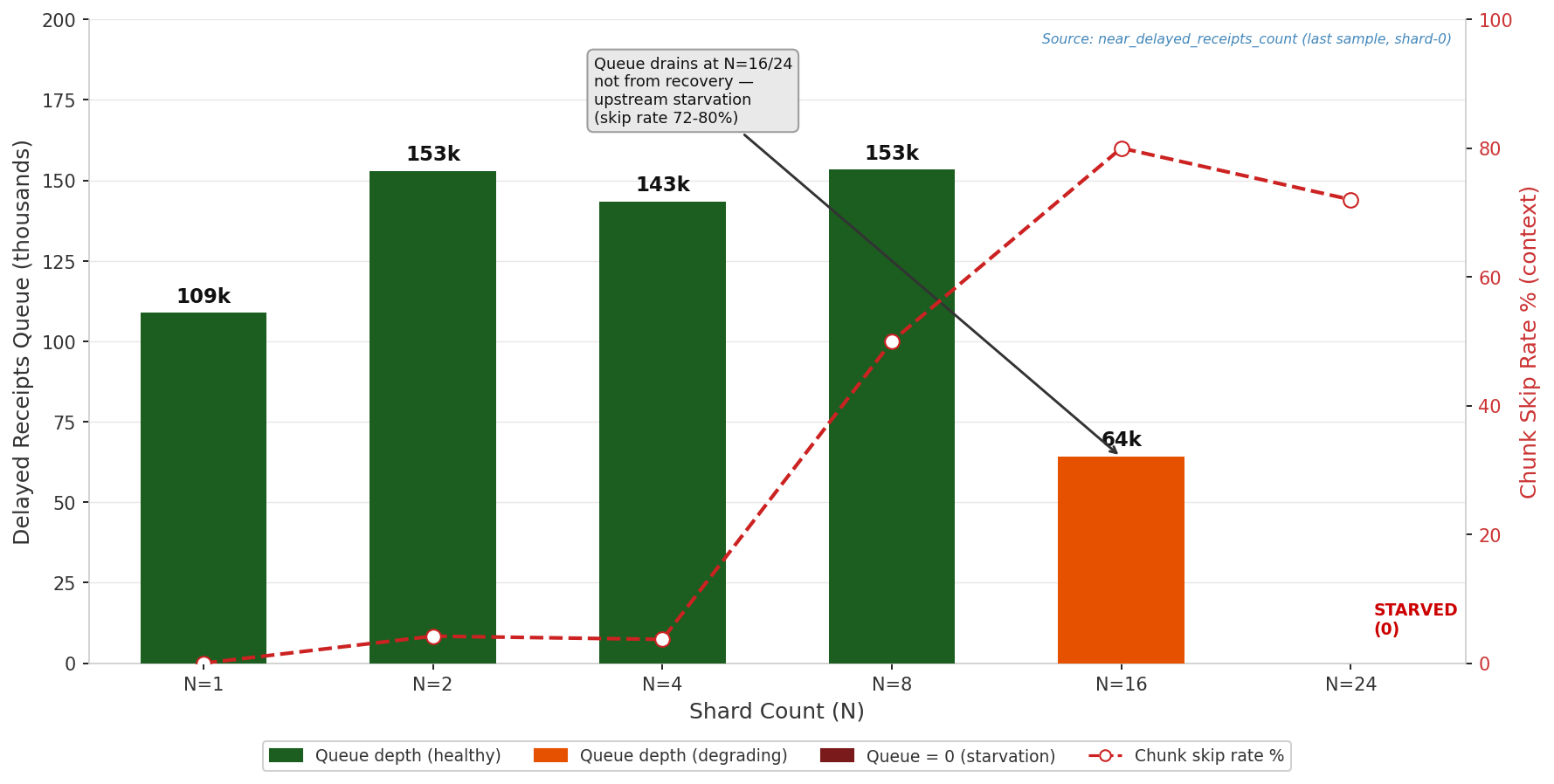}
  \caption{Delayed receipts queue depth vs shard count. The queue
    collapse at $N{=}16$ and $N{=}24$ is not recovery: it is upstream
    starvation caused by 72--80\% chunk skip rates.}
  \label{fig:receipts}
\end{figure}

The chunk skip rate is the most direct diagnostic of protocol health. At
$N{=}1$ through $N{=}4$, skip rate stays under 5\%. At $N{=}8$, skip
rate jumps to 50\%. By $N{=}16$, 80\% of chunk production attempts fail.
At $N{=}24$, the delayed receipts queue reaches zero: not from recovery
but from upstream starvation. The chain barely advances and gas\_used is
two orders of magnitude lower than all other $N$ values.

\subsection{HDD vs. tmpfs Comparison}

\begin{figure}[H]
  \centering
  \includegraphics[width=\columnwidth]{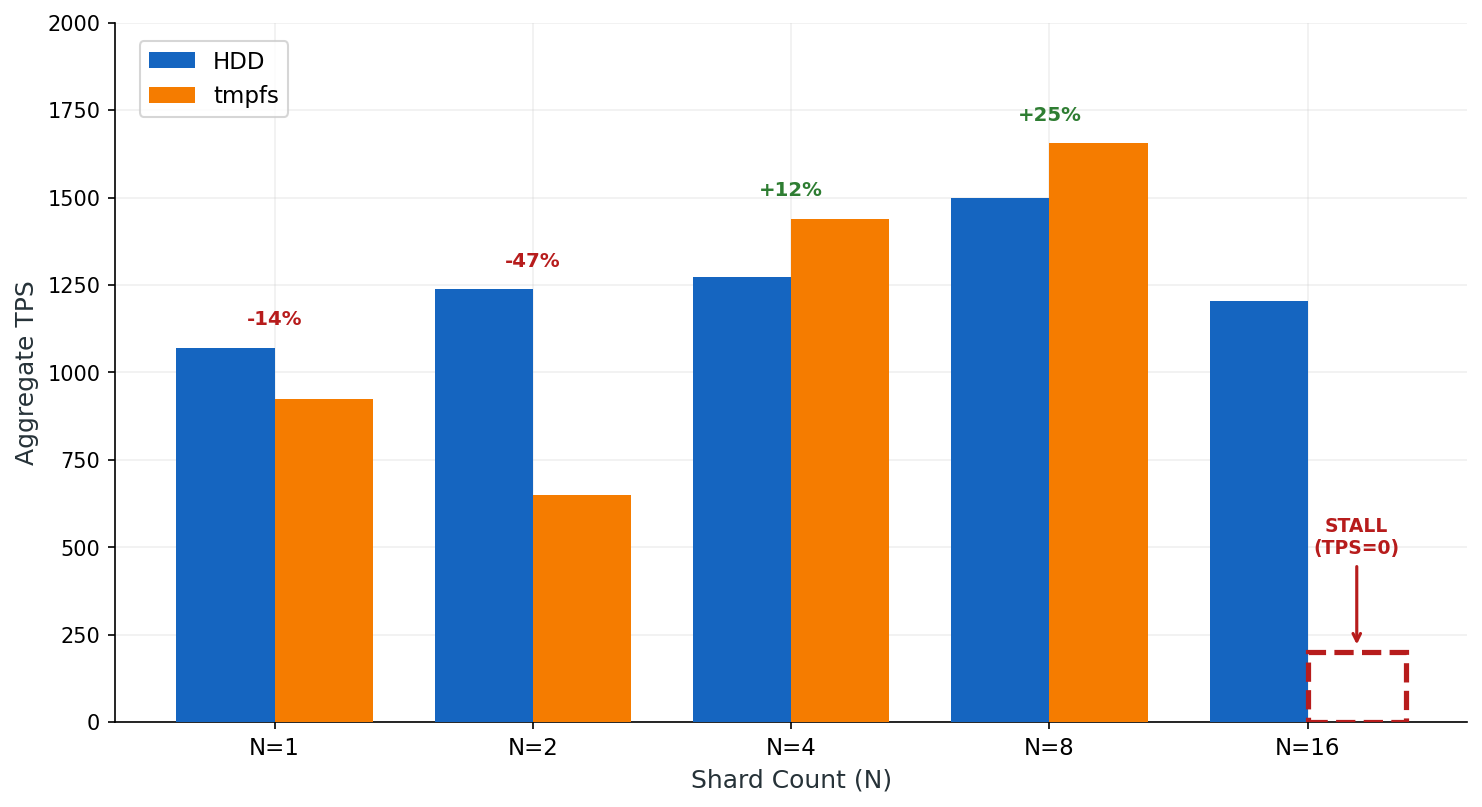}
  \caption{HDD vs tmpfs Aggregate TPS. At $N{=}16$ on tmpfs the chain
    stalled completely (TPS$=$0).}
  \label{fig:tmpfs}
\end{figure}

To test whether HDD throughput was the primary ceiling, RocksDB was moved
to \texttt{/dev/shm} (RAM-backed tmpfs), eliminating all disk I/O
(confirmed 0\,KB/s disk writes). Table~\ref{tab:tmpfs} shows the results.

\begin{table}[H]
\centering
\caption{HDD vs. tmpfs Aggregate TPS.}
\label{tab:tmpfs}
\renewcommand{\arraystretch}{1.15}
\begin{tabular}{C{0.4cm} C{0.9cm} C{0.9cm} C{0.65cm} L{2.5cm}}
\toprule
\textbf{N} & \textbf{HDD TPS} & \textbf{tmpfs TPS} & \textbf{Chg} &
\textbf{Key Observation} \\
\midrule
1  & 1,070 & 925   & $-$14\% & Page cache serving HDD reads \\
2  & 1,238 & 650   & $-$47\% & HDD pacing beneficial \\
4  & 1,272 & 1,440 & $+$12\% & Disk saturation removed \\
8  & 1,498 & 1,656 & $+$25\% & Gossip still ceiling \\
\textbf{16} & \textbf{1,204} & \textcolor{red}{\textbf{0 (STALL)}} &
  -- & Orphan witness 29$\times$; CPU 47\% \\
\bottomrule
\end{tabular}
\end{table}

Storage is not the primary bottleneck: tmpfs gave at most $+$25\% at
$N{=}8$, far below what a storage-bottlenecked system would show. The
$N{=}16$ chain stall is the critical finding. The HDD's write latency
paced chunk production to match finalization rate. Without it, witnesses
arrived before parent blocks finalized, orphan witness rate spiked 29$\times$,
and the chain halted. The protocol depended on disk I/O as implicit flow
control without knowing it.

\section{The Three Bottleneck Regimes}

\begin{figure}[H]
  \centering
  \includegraphics[width=\columnwidth]{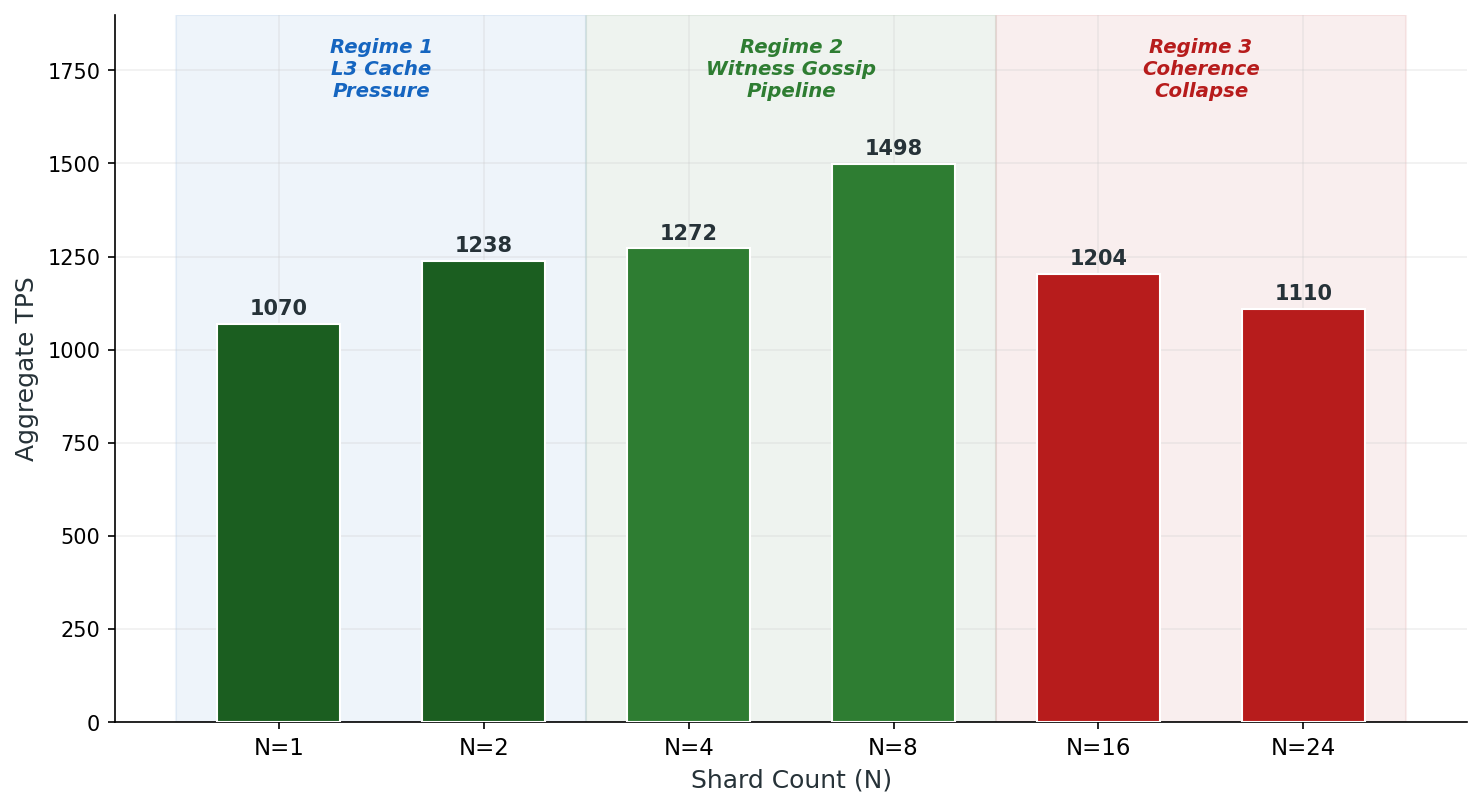}
  \caption{The three bottleneck regimes mapped onto aggregate TPS.
    Blue: L3 cache pressure ($N{=}1$--2). Green: witness gossip pipeline
    ($N{=}4$--8). Red: coherence collapse ($N{\geq}16$).}
  \label{fig:regimes}
\end{figure}

Fig.~\ref{fig:regimes} maps the three regimes onto aggregate TPS. TPS
peaks at the boundary between Regime 1 and Regime 2, where cache benefits
have been captured but gossip overhead has not yet dominated.

\textbf{Regime 1: L3 Cache Pressure ($N{=}1$--2).}
At low $N$, each validator traverses the full account trie per
transaction, causing LLC misses and DRAM pointer-chasing. The trie working
set overflows the $\sim$30--33\,MB L3 cache (LLC miss rate $\sim$15.7\%;
wall:ref ratio $>$5$\times$ at $N{=}1$). Sharding helps by partitioning
the trie across validators, shrinking each validator's working set and
improving cache hit rates. This cache benefit outweighs gossip cost until
$N{=}8$.

\textbf{Regime 2: Witness Gossip Pipeline ($N{=}4$--8).}
At mid-range $N$, the L3 cache benefit has been captured but the witness
gossip pipeline dominates. Each chunk requires encoding a $\sim$1\,MB
state witness, gossiping it to all $N$ validators, and awaiting
validation: approximately 600\,ms per chunk regardless of storage speed.
This is why tmpfs gives only modest gains at $N{=}4$--8. The
$\mathcal{O}(N^2)$ BFT approval messages add further overhead: at $N{=}8$,
28 unique validator pairs exchange 56 messages per block.

\textbf{Regime 3: Coherence Collapse ($N{\geq}16$).}
At high $N$, the witness gossip pipeline can no longer maintain the
invariant that witnesses arrive after their parent blocks finalize. With
HDD, disk write latency provides enough pacing to maintain coherence at
the edge (CV\%$=$27.6\%). Without it (tmpfs), witnesses arrive before
parents are finalized. Orphaned witnesses accumulate, trigger retry gossip,
generate a positive feedback loop, and drive TPS to zero. At $N{=}16$ on
tmpfs the chain stalls at 47\% CPU: the failure is coordination coherence,
not resource exhaustion.

\section{Key Findings}

\textbf{F1: Sharding peaks at $N{=}8$ then reverses.}
Aggregate TPS peaks at 1,498 at $N{=}8$ (+40\% over $N{=}1$). Beyond
$N{=}8$ it declines: 1,204 at $N{=}16$, 1,110 at $N{=}24$. Per-shard TPS
collapses 23$\times$ (1,070 to 46). Adding shards beyond $N{=}8$ actively
reduces total throughput on this hardware.

\textbf{F2: Bottleneck shifts from apply to gossip.}
At $N{=}1$, \texttt{apply\_chunks} is 1,118\,ms: 87\% of block processing
time. By $N{=}16$, apply is 179\,ms: only 4\% of 4,486\,ms total. The
4,307\,ms gap is witness gossip and BFT coordination, measured from
Prometheus \texttt{apply\_delay\_ms} $\Delta\text{sum}/\Delta\text{count}$.

\textbf{F3: Chunk skip rate collapses at high $N$.}
Skip rate is 0\% at $N{=}1$, 50\% at $N{=}8$, and 80\% at $N{=}16$.
At $N{=}24$ the skip rate is 72\% and the delayed receipts queue reaches
zero not from recovery but from upstream starvation: the chain barely
advances.

\textbf{F4: Storage is NOT the bottleneck.}
Replacing HDD with tmpfs (confirmed 0\,KB/s disk writes) improved TPS by
at most $+$25\% at $N{=}8$. The witness gossip pipeline is the real
ceiling at $N{\geq}4$, not storage.

\textbf{F5: HDD backpressure was load-bearing.}
At $N{=}16$ on tmpfs: TPS$=$0, CPU 47\%, orphan witness rate 29$\times$
higher than HDD. The HDD write latency paced chunk production to match
finalization rate. Without it, the chain entered coherence collapse. The
protocol depended on disk I/O as implicit flow control without knowing it.

\section{Implications and Future Work}

\textbf{For NEAR Protocol engineering.}
The witness gossip pipeline at $\sim$600\,ms per chunk at $N{\geq}4$ is
the primary optimization target. Reducing witness size through compression,
or reducing message count through tree-based aggregation
($\mathcal{O}(N \log N)$ instead of $\mathcal{O}(N^2)$), would directly
address the Regime 2 ceiling. The HDD pacing finding implies production
deployments rely on WAN latency for the same flow control role. This
co-tuning should be made explicit in the protocol, not left as an emergent
property of infrastructure.

\textbf{For the companion SimPy simulator~\cite{gandhi2026}.}
Our dataset provides the following calibration points: witness gossip time
$\sim$600\,ms per chunk at $N{=}4$--8 (loopback networking); apply time
decreasing from $\sim$880\,ms at $N{=}1$ to $\sim$50\,ms at $N{=}24$;
BFT overhead growing from $\sim$139\,ms to $\sim$2,555\,ms across the
full $N$ range; coherence collapse threshold at $N{=}16$ on single-node
tmpfs, approaching on HDD at $N{=}16$ (CV\%$=$27.6\%).

\textbf{Future work} includes: (i) the Phase~3 multi-node experiment
(4 validators on 4 dedicated Chameleon nodes) to confirm the memory bus
contention hypothesis and target 40,000--55,000 aggregate TPS (67--92\%
of the theoretical 59,200 TPS ceiling); (ii) WAN latency emulation via
\texttt{tc netem} 20--50\,ms to stabilize the $N{=}16$ tmpfs run;
(iii) cross-shard workloads with receipt processing; and (iv) witness
gossip profiling to decompose the $\sim$600\,ms into encoding,
transmission, and validation time.

\section{Conclusion}

We presented the first independent empirical characterization of NEAR
Protocol Nightshade sharding on commodity academic hardware. A systematic
sweep from $N{=}1$ to $N{=}24$ reveals that aggregate TPS peaks at
$N{=}8$ (+40\%) then reverses, while per-shard TPS collapses 23$\times$.
Block time grows 2.6$\times$ driven by $\mathcal{O}(N^2)$ BFT
coordination overhead. The tmpfs experiment produced the most surprising
finding: HDD write latency was implicitly providing the flow control that
the protocol depends on. The $N{=}16$ chain stall at 47\% CPU with a
29$\times$ spike in orphan witness rate demonstrates that the bottleneck
at high $N$ is coordination coherence, not hardware resources. The witness
gossip pipeline at $\sim$600\,ms per chunk at $N{\geq}4$ is the most
actionable target for protocol optimization. All code, data, and
instructions are publicly available~\cite{repo}.

\bibliographystyle{IEEEtran}
\bibliography{references}

\balance
\end{document}